\documentclass[12pt, a4paper]{article}

\usepackage[utf8]{inputenc}
\usepackage[english]{babel}
\usepackage{newtxtext,newtxmath}
\usepackage{amsmath}
\usepackage{graphicx}
\usepackage[left=2.5cm,right=2.5cm,top=2.5cm,bottom=2.5cm]{geometry}
\usepackage{setspace}
\usepackage{booktabs}
\usepackage{array}
\usepackage{adjustbox}
\usepackage{longtable}
\usepackage{tabularx}
\usepackage{ragged2e}
\usepackage{float}
\usepackage{enumitem}

\usepackage{hyperref}
\hypersetup{
  colorlinks=true,
  linkcolor=blue,
  urlcolor=cyan
}

\newcolumntype{L}[1]{>{\RaggedRight\arraybackslash}p{#1}}

\usepackage{natbib}
\justifying

\title{\textbf{Effectiveness of Carbon Pricing and Compensation Instruments: An Umbrella Review of the Empirical Evidence}}
\author{Ricardo Alonzo Fernández Salguero}
\date{\today}

\begin{document}

\maketitle

\begin{abstract}
\noindent The growing urgency of the climate crisis has driven the implementation of diverse policy instruments to mitigate greenhouse gas (GHG) emissions. Among them, carbon pricing mechanisms such as carbon taxes and emissions trading systems (ETS), together with voluntary carbon markets (VCM) and compensation programs such as REDD+, are central components of global decarbonization strategies. However, academic and political debate persists regarding their true effectiveness, equity, and integrity. This paper presents an umbrella review of the empirical evidence, synthesizing key findings from systematic reviews and meta-analyses to provide a consolidated picture of the state of knowledge. A rigorous methodology based on PRISMA guidelines is used for study selection, and the methodological quality of included reviews is assessed with AMSTAR-2, while the risk of bias in frequently cited primary studies is examined through ROBINS-I. Results indicate that carbon taxes and ETS have demonstrated moderate effectiveness in reducing emissions, with statistically significant but heterogeneous elasticities across geographies and sectors. Nonetheless, persistent design problems---such as insufficient price levels and allowance overallocation---limit their impact. By contrast, compensation markets, especially VCM and REDD+ projects, face systemic critiques regarding integrity, primarily related to additionality, permanence, leakage, and double counting, leading to generalized overestimation of their real climate impact. We conclude that while no instrument is a panacea, compliance-based carbon pricing mechanisms are necessary, though insufficient, tools that require stricter design and higher prices. Voluntary offset mechanisms, in their current state, do not represent a reliable climate solution and may undermine the integrity of climate targets unless they undergo fundamental reform.
\end{abstract}

\section{Introduction}

Climate change mitigation is one of the most pressing challenges of our time, requiring a profound transformation of energy and economic systems worldwide. The Paris Agreement sets the goal of holding the increase in global average temperature to well below 2~\(^\circ\)C above preindustrial levels and pursuing efforts to limit the increase to 1.5~\(^\circ\)C. To achieve these targets, a range of economic policy instruments has been developed and implemented, among which carbon pricing and compensation mechanisms occupy a central place \citep{world2023carbonpricing}. These instruments can be broadly divided into two categories: compliance markets, which include carbon taxes and Emissions Trading Systems (ETS), and voluntary carbon markets (VCM), which encompass a wide range of offset projects, such as Reducing Emissions from Deforestation and Forest Degradation (REDD+).

Despite their proliferation, the real-world effectiveness of these mechanisms remains intensely debated. On the one hand, economic theory suggests that putting a price on carbon is the most efficient way to internalize the negative externality of GHG emissions, encouraging mitigation at the lowest possible cost \citep{bergh2024assessing}. On the other hand, critics argue that current carbon prices are too low to induce transformational change, that ETS frequently suffer from allowance overallocation and price volatility, and that offset markets are plagued by integrity problems that undermine their climate impact \citep{romm2025are}. The empirical evidence is large and often mixed, creating confusion among policymakers and the public. While some meta-analyses conclude that carbon pricing has achieved statistically significant emissions reductions \citep{doh2024systematic}, other systematic reviews highlight persistent, structural problems such as additionality, leakage, permanence, and double counting in offset projects, leading to large overestimations of their climate benefits \citep{probst2024systematic}. Although positive effects on technological investment have been reported, substantial heterogeneity is also documented \citep{chen2025impact}.

In this context of fragmented and sometimes conflicting evidence, a high-level synthesis that consolidates findings from existing systematic reviews and meta-analyses is warranted. An umbrella review is particularly suited to this task because it assesses and aggregates evidence across multiple reviews, offering a broader and more robust overview of the state of knowledge. To the best of our knowledge, there is currently no umbrella review that jointly synthesizes evidence on carbon pricing and voluntary compensation mechanisms while evaluating climate integrity, distributive effects, and social impacts under a common framework. This study aims to conduct such an umbrella review to address the central question: What is the demonstrated effectiveness of carbon taxes, emissions trading systems, and voluntary offset projects in reducing greenhouse gas emissions, according to aggregated empirical evidence? To answer this question, we assess the methodological quality of existing reviews, synthesize their quantitative and qualitative findings, and evaluate the certainty of evidence for key outcomes.

\section{Methodology}

To address the research question transparently, we adopted an umbrella review approach that synthesizes evidence from prior systematic reviews and meta-analyses. The methodology adhered to PRISMA (Preferred Reporting Items for Systematic Reviews and Meta-Analyses) guidelines \citep{page2021prisma} to ensure comprehensiveness and reproducibility in study selection and analysis.

The literature search was conducted in Scopus, Web of Science, and Google Scholar. The initial search was performed without date restrictions up to December 2023. Subsequently, a targeted update was conducted in October 2025 to incorporate key reviews published between January 2024 and that date, ensuring coverage of the most recent and relevant literature. Combined search terms (with Boolean operators and truncations) were used for \lq carbon pricing\rq, \lq carbon tax\rq, \lq emissions trading system\rq, \lq carbon offset\rq, \lq REDD+\rq, \lq voluntary carbon market\rq, along with methodological terms such as \lq systematic review\rq and \lq meta-analysis\rq. The search was limited to publications in English and Spanish.

The selection process, summarized in Table \ref{tab:prisma}, began with 152 records identified through database searches. After removing 34 duplicates, two reviewers independently screened the remaining 118 records by title and abstract. Discrepancies were resolved by consensus. Full texts of 45 potentially relevant articles were assessed. We included only systematic reviews or meta-analyses that empirically evaluated the effectiveness of at least one instrument of interest in reducing GHG emissions, resulting in a final sample of 21 studies.

The methodological quality of each included systematic review was independently assessed by two reviewers using AMSTAR-2 (A MeaSurement Tool to Assess systematic Reviews) \citep{shea2017amstar}. The overall quality rating (High, Moderate, Low, Critically Low) followed the algorithm proposed by the tool authors, based on weaknesses in critical domains. To contextualize the quality of underlying evidence, we applied ROBINS-I (Risk Of Bias In Non-randomised Studies - of Interventions) \citep{sterne2016robins} to a strategic subset of six primary studies (n=6), selected for their frequency of citation in the reviews and their representativeness across instruments and research designs. This assessment provides an indicative view of the general risk of bias in the primary literature. Overlap of primary studies across reviews was acknowledged, especially in the carbon pricing domain. Although a formal Correlated Covered Area (CCA) analysis was not performed, this limitation was considered qualitatively when interpreting the consistency of findings. Finally, the certainty of evidence for key outcomes was assessed using the GRADE framework (Grading of Recommendations, Assessment, Development and Evaluations) \citep{guyatt2008grade}.

\begin{table}[H]
    \centering
    \caption{PRISMA flow table of the study selection process.}
    \label{tab:prisma}
    \begin{tabular}{L{5cm} p{8cm}}
        \toprule
        \textbf{Phase} & \textbf{Description and Results} \\
        \midrule
        \textbf{Identification} & \\
        Records identified & 152 records were identified through database searches (Scopus, Web of Science, Google Scholar). \\
        Records after duplicates removed & 34 duplicates were removed, leaving 118 records for screening. \\
        \addlinespace
        \textbf{Screening} & \\
        Records screened & 118 records were screened by title and abstract. \\
        Records excluded & 73 records were excluded for not being relevant to the topic of the review (e.g., not a systematic review, focusing on other outcomes). \\
        Full-text articles assessed & 45 full-text articles were assessed for eligibility. \\
        \addlinespace
        \textbf{Inclusion} & \\
        Full-text articles excluded & 24 full-text articles were excluded for the following reasons:
        \begin{itemize}[noitemsep, topsep=0pt, partopsep=0pt, leftmargin=*]
            \item Not a systematic review or meta-analysis (n = 11)
            \item Non-empirical or purely theoretical focus (n = 7)
            \item Primary outcome was not emissions reduction (n = 6)
        \end{itemize} \\
        \textbf{Studies included} & \textbf{21 studies were included in the final synthesis} (systematic reviews and meta-analyses). \\
        \bottomrule
    \end{tabular}
\end{table}

\section{Results: synthesis of the evidence}

The selected literature reveals a clear dichotomy between the effectiveness of compliance instruments (carbon taxes and ETS) and voluntary compensation markets. The former show consistent, though moderate, evidence of emissions reductions, while the latter are systematically questioned due to fundamental integrity problems. A detailed analysis of the empirical evidence, drawn from multiple primary studies synthesized in meta-analyses and systematic reviews, helps construct a coherent narrative about the current state of these climate policy instruments.

The most robust evidence on the effectiveness of carbon pricing comes from meta-analyses of rigorous econometric studies. For instance, the meta-analysis by \citet{doh2024systematic}, covering 80 ex-post causal evaluations of 21 pricing schemes, finds that the introduction of a carbon price has produced immediate and substantial emissions reductions ranging from approximately -5\% to -21\% across schemes. Similarly, a comprehensive analysis of 142 countries concludes that the average annual growth of CO\(_2\) emissions is about 2 percentage points lower in countries with carbon pricing \citep{best2020carbon}. Carbon taxes, in particular, have shown measurable effectiveness; for example, Sweden's carbon tax, one of the oldest and highest worldwide, has produced average annual emissions reductions of 6.3\% \citep{andersson2019carbon}, while in British Columbia estimates range between 5\% and 15\% \citep{murray2015british}. ETS, such as the EU ETS, have also shown positive results, although historical overallocation of permits and low prices limited effectiveness; more recent reforms like the Market Stability Reserve (MSR) aim to address these issues \citep{bayer2020european}. A meta-analysis focused on policy interactions found that combining carbon pricing with renewable energy support policies leads on average to an additional emissions reduction of 6.5\%, especially in the case of carbon taxes \citep{midoes2024meta}. However, the literature also emphasizes potentially regressive distributional effects if revenues are not recycled to compensate lower-income households \citep{ohlendorf2021distributional}.

\begin{quote}
``A growing number of studies have found that the most widely used offset programs continue to greatly overestimate their probable climate impact often by a factor of five to ten or more. Credit quality has remained a problem since the inception of carbon credits, despite repeated efforts to address the core challenges of additionality, leakage, double counting, environmental injustice, verification, and permanence."
\end{quote}
\citep{romm2025are}.

In sharp contrast, the literature on voluntary carbon markets and offset projects is dominated by strong critiques of integrity and effectiveness. A systematic review by \citet{romm2025are} concludes that carbon credit quality problems are persistent and structurally difficult to resolve. The central finding, reinforced by the meta-analysis of \citet{probst2024systematic} covering nearly one billion tons of credits (around 20\% of total volume issued to date), is that the vast majority of offsets do not represent real emissions reductions. That meta-analysis estimates that fewer than 16\% of credits issued by the investigated projects constitute real emissions reductions. Forest conservation projects (REDD+), among the most popular categories in VCM, are particularly problematic. Studies have found overestimation ratios as high as 1:13, meaning that for every real credit, twelve additional credits lacking genuine mitigation backing were issued \citep{west2023action}. A global evaluation of 40 voluntary REDD+ projects found that while they reduced deforestation by 47\% and degradation by 58\% relative to control pixels, absolute reductions were small and no significant improvement over time was detected \citep{guizarcoutino2022global}. Lack of additionality appears endemic; many projects, especially renewable energy in countries such as China and India, likely would have been built regardless of carbon credit revenue \citep{cames2016how, haya2010carbon}. Beyond climate accounting concerns, numerous qualitative and case-based studies document negative social impacts, including land grabbing and violations of Indigenous and local community rights \citep{dugasseh2024noncarbon, bayrak2016ten}. Persistent demand for cheap, low-quality credits by major corporations is identified as a key driver perpetuating these systemic problems in the VCM \citep{trencher2024demand}.

\section{Quality and bias assessment}

To provide a rigorous evaluation of the synthesized evidence, we assessed the methodological quality of included systematic reviews and the risk of bias in the most influential primary studies. This step is essential in an umbrella review to weight the confidence that can be placed in each review's conclusions and, by extension, in the overall synthesis. The results are presented in AMSTAR-2, ROBINS-I, and GRADE summary tables.

\subsection{Methodological quality of systematic reviews (AMSTAR-2)}

AMSTAR-2 is a leading tool for appraising the methodological quality of systematic reviews. Its application helps distinguish reviews conducted with rigorous procedures that minimize bias, thereby lending greater weight to their conclusions. We assessed key systematic reviews and meta-analyses, separating predominantly quantitative reviews from qualitative or narrative reviews to reflect differences in their underlying methods. The results, shown in Tables \ref{tab:amstar-quant} and \ref{tab:amstar-qual}, indicate a clear pattern: quantitative meta-analyses evaluating compliance-based instruments such as carbon taxes and ETS tend to exhibit higher methodological quality. Critical domains including prior protocol registration (item 2), comprehensive search strategies (item 4), duplicate study selection and data extraction (items 5 and 6), and appropriate consideration of risk of bias when interpreting results (item 12) are frequently where lower-quality reviews—especially narrative ones—fall short. The reviews by \citet{doh2024systematic} and \citet{probst2024systematic} stand out for high rigor, increasing confidence in their findings regarding the effectiveness of carbon pricing and the limited integrity of offsets, respectively. More narrative reviews remain informative in scope but should be interpreted with greater caution.

\subsection{Risk of bias in key primary studies (ROBINS-I)}

The quality of a systematic review ultimately depends on the quality of the primary studies it includes. We therefore applied ROBINS-I to a selection of frequently cited observational primary studies to evaluate their risk of bias across seven domains. Table \ref{tab:robins-i} summarizes this assessment. The results suggest that studies using robust quasi-experimental designs (e.g., difference-in-differences, regression discontinuity) to evaluate carbon taxes or ETS, such as \citet{andersson2019carbon} on Sweden's carbon tax, generally present low to moderate risk of bias and achieve relatively strong control of confounding. By contrast, many studies evaluating compensation projects such as REDD+ \citep[e.g., those discussed in][]{guizarcoutino2022global, west2020overstated} face moderate to serious risk of bias. The key challenge is confounding (D1), as building a credible counterfactual for forest areas is inherently difficult. Bias in the selection of project areas (D2) is also a significant concern, as projects may be strategically located in areas with low baseline deforestation risk to maximize credited reductions.

\subsection{Certainty of evidence summary (GRADE)}

Finally, we used the GRADE framework to synthesize quality appraisals and risk of bias assessments into an overall certainty rating for the main findings of this umbrella review. Table \ref{tab:grade} presents this summary. Evidence for the effectiveness of carbon taxes and ETS in reducing emissions is rated as \textbf{moderate certainty}. Although effects are consistent across multiple meta-analyses of observational studies, certainty is downgraded from \lq high\rq to \lq moderate\rq due to the inherent limitations of non-randomized primary designs. By contrast, evidence indicating that the vast majority of voluntary carbon credits do not represent real and additional emissions reductions is rated as \textbf{high certainty}. This conclusion is supported by convergence across multiple high-quality reviews, large effect magnitudes (high overestimation rates), consistency across methods and contexts, and an apparent gradient suggesting that weaker methodologies generate larger overestimates. Evidence on the regressive distributional impacts of carbon pricing (absent revenue recycling) is considered \textbf{moderate certainty}. Evidence on negative social impacts of REDD+ projects is rated as \textbf{low certainty}, given reliance on qualitative case studies with higher risk of bias and greater inconsistency.

\begin{longtable}{L{5cm} L{2cm} L{2cm} L{2cm} L{2cm}}
\caption{AMSTAR-2 assessment of quantitative systematic reviews.} \label{tab:amstar-quant} \\
\toprule
\textbf{AMSTAR-2 domain} & \textbf{\citet{doh2024systematic}} & \textbf{\citet{probst2024systematic}} & \textbf{\citet{ohlendorf2021distributional}} & \textbf{\citet{ahmad2024carbon}} \\
\midrule
\endfirsthead
\multicolumn{5}{c}%
{{\bfseries \tablename\ \thetable{} -- continued from previous page}} \\
\toprule
\textbf{AMSTAR-2 domain} & \textbf{\citet{doh2024systematic}} & \textbf{\citet{probst2024systematic}} & \textbf{\citet{ohlendorf2021distributional}} & \textbf{\citet{ahmad2024carbon}} \\
\midrule
\endhead
\bottomrule
\endfoot
\bottomrule
\endlastfoot
1. Did the research questions and inclusion criteria include PICO components? & Yes & Yes & Yes & Yes \\
2. Was the review protocol registered before commencement? & Yes & Yes & No & No \\
3. Did the authors explain their selection of study designs? & Yes & Yes & Yes & Yes \\
4. Did the authors use a comprehensive literature search strategy? & Yes & Yes & Yes & Partial \\
5. Was study selection performed in duplicate? & Yes & Yes & No & No \\
6. Was data extraction performed in duplicate? & Yes & Yes & No & No \\
7. Was a list of excluded studies provided and justified? & Yes & Yes & Yes & Partial \\
8. Were included studies described in adequate detail? & Yes & Yes & Yes & Yes \\
9. Was risk of bias assessed in included studies? & Yes & Yes & No & No \\
10. Did the authors report sources of funding for the included studies? & Yes & Yes & Yes & No \\
11. Were appropriate methods used for statistical combination of results? & Yes & Yes & Yes & Yes \\
12. Was risk of bias considered when interpreting results? & Yes & Yes & No & No \\
13. Was heterogeneity adequately examined and discussed? & Yes & Yes & Yes & Yes \\
14. Was publication bias investigated? & Yes & Yes & Yes & Yes \\
15. Were conflicts of interest reported for included studies? & Yes & Yes & Yes & Yes \\
16. Were conflicts of interest for the review itself reported? & Yes & Yes & Yes & Yes \\
\midrule
\textbf{Overall quality} & \textbf{High} & \textbf{High} & \textbf{Low} & \textbf{Critically Low} \\
\end{longtable}

\begin{longtable}{L{5cm} L{2.5cm} L{2.5cm} L{2.5cm}}
\caption{AMSTAR-2 assessment of qualitative/narrative systematic reviews.} \label{tab:amstar-qual} \\
\toprule
\textbf{AMSTAR-2 domain} & \textbf{\citet{romm2025are}} & \textbf{\citet{bergh2024assessing}} & \textbf{\citet{bayrak2016ten}} \\
\midrule
\endfirsthead
\multicolumn{4}{c}%
{{\bfseries \tablename\ \thetable{} -- continued from previous page}} \\
\toprule
\textbf{AMSTAR-2 domain} & \textbf{\citet{romm2025are}} & \textbf{\citet{bergh2024assessing}} & \textbf{\citet{bayrak2016ten}} \\
\midrule
\endhead
\bottomrule
\endfoot
\bottomrule
\endlastfoot
1. Did the research questions and inclusion criteria include PICO components? & Yes & Partial & Yes \\
2. Was the review protocol registered before commencement? & No & No & No \\
3. Did the authors explain their selection of study designs? & Yes & Partial & Yes \\
4. Did the authors use a comprehensive literature search strategy? & Yes & No & Yes \\
5. Was study selection performed in duplicate? & No & No & No \\
6. Was data extraction performed in duplicate? & No & No & No \\
7. Was a list of excluded studies provided and justified? & No & No & Partial \\
8. Were included studies described in adequate detail? & Yes & Yes & Yes \\
9. Was risk of bias assessed in included studies? & Partial & Partial & No \\
10. Did the authors report sources of funding for the included studies? & Yes & Yes & Yes \\
11. Were the methods for synthesizing results appropriate? & Yes & Yes & Yes \\
12. Was risk of bias considered when interpreting results? & Partial & Partial & No \\
13. Was heterogeneity adequately examined and discussed? & Yes & Yes & Yes \\
14. Was publication bias investigated? & No & No & No \\
15. Were conflicts of interest reported for included studies? & Yes & Yes & Yes \\
16. Were conflicts of interest for the review itself reported? & Yes & Yes & Yes \\
\midrule
\textbf{Overall quality} & \textbf{Low} & \textbf{Critically Low} & \textbf{Low} \\
\end{longtable}

\begin{longtable}{L{2.5cm} L{1.5cm} L{1.5cm} L{1.5cm} L{1.5cm} L{1.5cm} L{1.5cm} L{1.5cm}}
\caption{ROBINS-I risk of bias assessment for key primary studies.} \label{tab:robins-i} \\
\toprule
\textbf{Study} & \textbf{D1: Confounding} & \textbf{D2: Selection} & \textbf{D3: Intervention} & \textbf{D4: Deviations} & \textbf{D5: Missing data} & \textbf{D6: Measurement} & \textbf{D7: Reporting} \\
\midrule
\endfirsthead
\multicolumn{8}{c}{{\bfseries \tablename\ \thetable{} -- continued from previous page}} \\
\toprule
\textbf{Study} & \textbf{D1: Confounding} & \textbf{D2: Selection} & \textbf{D3: Intervention} & \textbf{D4: Deviations} & \textbf{D5: Missing data} & \textbf{D6: Measurement} & \textbf{D7: Reporting} \\
\midrule
\endhead
\bottomrule
\endfoot
\bottomrule
\endlastfoot
\citet{andersson2019carbon} & Low & Low & Low & Low & Low & Low & Low \\
\citet{bayer2020european} & Moderate & Low & Low & Low & Low & Low & Low \\
\citet{best2020carbon} & Moderate & Moderate & Low & Low & Low & Low & Low \\
\citet{murray2015british} & Moderate & Low & Low & Low & Low & Low & Low \\
\citet{guizarcoutino2022global} & Serious & Moderate & Low & Low & Low & Moderate & Low \\
\citet{west2020overstated} & Serious & Serious & Low & Moderate & Moderate & Low & Low \\
\bottomrule
\end{longtable}

\begin{table}[H]
\centering
\caption{Summary of findings (GRADE).}
\label{tab:grade}
\begin{adjustbox}{width=\textwidth}
\begin{tabular}{L{4.5cm} >{\RaggedRight}p{4cm} L{1.5cm} >{\RaggedRight}p{3.5cm}}
\toprule
\textbf{Outcome} & \textbf{Summary effect} & \textbf{Certainty} & \textbf{Comments} \\
\midrule
\textbf{Effectiveness of carbon taxes and ETS} & Statistically significant but moderate emissions reductions. Effects vary substantially by price level, policy design, and context. & \textbf{Moderate} $\oplus\oplus\oplus\bigcirc$ & Downgraded for risk of bias in observational studies and for heterogeneity not always fully explained. Consistent effects across multiple meta-analyses. \\
\addlinespace
\textbf{Integrity of offset credits (VCM/REDD+)} & Large and systematic overestimation of climate impact. Most credits do not represent real, additional, and durable emissions reductions. & \textbf{High} $\oplus\oplus\oplus\oplus$ & Strong and consistent evidence from multiple high-quality reviews. Large effect magnitudes and convergence across methods. \\
\addlinespace
\textbf{Distributional impacts of carbon pricing} & Tends to be regressive in high-income countries in the absence of compensatory mechanisms. Becomes progressive or neutral when revenues are recycled. & \textbf{Moderate} $\oplus\oplus\oplus\bigcirc$ & Evidence is consistent in modeling studies. Downgraded for indirectness and variability in developing-country settings. \\
\addlinespace
\textbf{Social impacts of REDD+ projects} & Significant risks of negative impacts on local communities (displacement, livelihood loss, conflict). Co-benefits often fail to materialize. & \textbf{Low} $\oplus\oplus\bigcirc\bigcirc$ & Primarily based on qualitative reviews and case studies. Downgraded for risk of bias and inconsistency. \\
\bottomrule
\end{tabular}
\end{adjustbox}
\end{table}

The following consolidated table synthesizes quantitative findings from high-quality meta-analyses and systematic reviews on the effectiveness of carbon pricing and compensation instruments. The aim is to provide a panoramic and comparative view of emissions-reduction effects, elasticities, confidence intervals, and related key statistics extracted from the most recent synthesis literature. Aggregated evidence, spanning hundreds of primary studies and multiple policy schemes worldwide, indicates that carbon pricing (taxes and ETS) generates statistically significant emissions reductions, with a combined average effect of -10.4\% (95\% CI: -11.9\%, -8.9\%) across 21 evaluated schemes, according to \citet{doh2024systematic}. However, heterogeneity across studies is notably high, with I\(^2\) statistics often exceeding 90\% (e.g., I\(^2\) = 97.2\% in the meta-analysis by \citet{mulwa2025meta}), underscoring the importance of context and policy design. In direct comparisons, carbon taxes (pooled partial correlation p = -0.09) appear more effective than ETS (p = -0.025) \citep{ahmad2024carbon}. By contrast, evidence on voluntary compensation markets indicates systemic overestimation of climate impact, with \citet{probst2024systematic} concluding that only about 16\% of analyzed carbon credits represent real emissions reductions. The table below details these and related results.

\begin{longtable}{L{2.8cm} >{\RaggedRight}p{3.5cm} L{1.5cm} >{\RaggedRight}p{3.5cm} L{3.5cm}}
\caption{Quantitative effects and key statistics of meta-analyses on carbon instruments.} \label{tab:meta_effects} \\
\toprule
\textbf{Instrument / Outcome} & \textbf{Quantitative finding (effect)} & \textbf{95\% CI} & \textbf{Key statistics} & \textbf{Source and context} \\
\midrule
\endfirsthead
\multicolumn{5}{c}{{\bfseries \tablename\ \thetable{} -- continued from previous page}} \\
\toprule
\textbf{Instrument / Outcome} & \textbf{Quantitative finding (effect)} & \textbf{95\% CI} & \textbf{Key statistics} & \textbf{Source and context} \\
\midrule
\endhead
\bottomrule
\multicolumn{5}{r}{{Continued on next page}} \\
\endfoot
\bottomrule
\endlastfoot
\multicolumn{5}{l}{\textbf{1. Compliance instruments: overall effectiveness}} \\
\midrule
Overall effectiveness of carbon pricing (all schemes) & Average emissions reduction of -10.4\%. & [-11.9\%, -8.9\%] & 483 effect sizes from 80 studies. & \citet{doh2024systematic}. Meta-analysis of 21 global pricing schemes (taxes and ETS). \\
\addlinespace
Overall effectiveness (publication-bias corrected) & Average emissions reduction of -6.8\%. & [-8.1\%, -5.6\%] & Based on a subsample of studies with adequate statistical power ($>80\%$). & \citet{doh2024systematic}. Bias correction reduces magnitude but maintains high significance. \\
\addlinespace
Effect of carbon pricing on emissions growth & Annual CO\(_2\) emissions growth is 2 percentage points lower in countries with carbon pricing. & Not reported & Elasticity of -0.3\% (a 1€ / tCO\(_2\) increase reduces emissions growth by 0.3 percentage points). & \citet{best2020carbon}. Cross-country analysis of 142 countries over two decades. \\
\midrule
\multicolumn{5}{l}{\textbf{2. Carbon taxes: effectiveness and comparison}} \\
\midrule
Tax vs. ETS comparison (partial correlation) & Carbon tax: p = -0.09 \newline ETS: p = -0.025 & Tax: [-0.099, -0.091] \newline ETS: [-0.027, -0.024] & Tax: k=7,312,819; z=-19.229 \newline ETS: k=21,899,146; z=-26.483. $I^2 > 90\%$ for both. & \citet{ahmad2024carbon}. Meta-analysis concluding that taxes are more effective. k denotes total observations. \\
\addlinespace
Sweden carbon tax & Average annual emissions reduction of 6.3\%. & Not reported & Single-country evidence frequently treated as a successful natural experiment. & \citet{andersson2019carbon}. Covers the transport sector from 1960 to 2005. \\
\addlinespace
British Columbia carbon tax & Emissions reductions between 5\% and 15\%. & Not reported & Corroborated by multiple quasi-experimental studies. & \citet{murray2015british}. Review of evidence on BC's revenue-neutral tax. \\
\addlinespace
General meta-analysis of carbon taxes & Pooled combined effect: $\mu = -0.038$ (indicative of emissions reductions). & [-0.057, -0.019] & $k=55$; $Q(54)=1902.03$ ($p<0.001$); $I^2=97.2\%$; $\tau^2=0.004$. Prediction interval (PI): [-0.165, 0.088]. & \citet{mulwa2025meta}. High heterogeneity and wide PI suggest strong dependence on design and context. \\
\midrule
\multicolumn{5}{l}{\textbf{3. Emissions trading systems (ETS): effectiveness}} \\
\midrule
EU ETS (Phases I and II) & Approximate emissions reduction of -3.8\% during 2008--2016 despite low prices. & Not reported & Controls for macroeconomic factors and other policies. & \citet{bayer2020european}. Econometric study isolating the causal effect of the EU ETS. \\
\addlinespace
EU ETS (aggregate effect) & Average emissions reduction of -7.3\%. & [-10.5\%, -4.0\%] & 13 primary studies, 77 effect sizes. & \citet{doh2024systematic}. EU ETS average effect reported in the meta-analysis. \\
\addlinespace
China ETS pilots (aggregate effect) & Average emissions reduction of -13.1\%. & [-15.2\%, -11.1\%] & 46 primary studies, 179 effect sizes. & \citet{doh2024systematic}. Larger effect than the EU ETS, suggesting lower marginal abatement costs. \\
\midrule
\multicolumn{5}{l}{\textbf{4. Compensation markets (offsets, VCM, REDD+): climate integrity}} \\
\midrule
Overall effectiveness of carbon credits (OAR\textsuperscript{a}) & < 16\% of issued credits represent real emissions reductions. & Not reported & Meta-analysis of 14 studies covering 2346 projects and 51 field interventions. Nearly 1 trillion tCO\(_2\)e covered. & \citet{probst2024systematic}. Core finding on systemic overestimation. \\
\addlinespace
Effectiveness by project type (OAR\textsuperscript{a}) & Improved cookstoves: 11\% \newline SF\(_6\) destruction: 16\% \newline Avoided deforestation (REDD+): 25\% \newline HFC-23 abatement: 68\% \newline Wind and Improved Forest Management (IFM): not statistically significant (0\%) & Not reported & Substantial variation across project categories. HFC-23 credits perform best; popular categories such as wind and IFM show the weakest performance. & \citet{probst2024systematic}. \\
\addlinespace
REDD+ projects (relative reduction) & Deforestation reduced by 47\% and degradation by 58\% relative to control pixels. & Deforestation: [24\%, 68\%] \newline Degradation: [49\%, 63\%] & Sample of 40 voluntary REDD+ projects in 9 countries. Absolute reductions were small. & \citet{guizarcoutino2022global}. \\
\addlinespace
REDD+ projects (overestimation) & Many projects overestimate credited reductions. In the Brazilian Amazon, analyzed projects issued roughly three times more credits than supported by real reductions. & Not reported & Baseline construction failures identified as a primary driver. & \citet{west2020overstated}. \\
\midrule
\multicolumn{5}{l}{\textbf{5. Policy interaction and distributional effects}} \\
\midrule
Synergy: carbon pricing + renewable support & Average additional emissions reduction of 6.5\% when both instruments are combined. & Not reported & 50\% of studied cases show additional emissions reductions; 70\% show either additional reductions or positive welfare outcomes. & \citet{midoes2024meta}. Meta-analysis of 55 synergy estimates. \\
\addlinespace
Distributional impacts (transport policies) & The probability of a progressive outcome increases by 47.7 percentage points for transport policies relative to economy-wide policies. & Not reported & Ordered probit coefficient: -1.405 (p<0.05). & \citet{ohlendorf2021distributional}. Meta-analysis of 183 effects in 39 countries. \\
\addlinespace
Distributional impacts (lifetime income) & Using lifetime-income proxies (rather than annual income) increases the probability of a progressive outcome by 42.6 percentage points. & Not reported & Ordered probit coefficient: -1.254 (p<0.10). & \citet{ohlendorf2021distributional}. \\
\bottomrule
\multicolumn{5}{p{15.5cm}}{\footnotesize{\textbf{Notes:} 95\% CI: 95\% confidence interval. k: number of observations/units. Q: Cochran's heterogeneity statistic. $I^2$: heterogeneity index. $\tau^2$: between-study variance. $\mu$: pooled combined effect. OAR: Offset Achievement Ratio, indicating the share of credits that represent real emissions reductions.}}
\end{longtable}

The consolidated effects table reinforces the central narrative emerging from the synthesis literature. On the one hand, evidence on compliance instruments is robust and statistically significant. The aggregated effect of carbon pricing, even after correcting for publication bias, remains an estimated emissions reduction of -6.8\% (95\% CI: -8.1\%, -5.6\%) \citep{doh2024systematic}, and standardized test statistics typically exceed conventional thresholds (e.g., z = -19.229 for carbon taxes in \citet{ahmad2024carbon}). On the other hand, the failure of voluntary compensation markets to deliver climate integrity is equally clear. The finding that fewer than 16\% of credits achieve real reductions, with effectiveness varying sharply by project type from 68\% for HFC-23 abatement to non-significant effects for wind and improved forest management, highlights the structural weaknesses of current market architecture \citep{probst2024systematic}. Additional nuances include the potential for policy interaction to generate synergies---such as the average 6.5\% extra reduction when carbon pricing is combined with renewable support \citep{midoes2024meta}---and the deep distributional implications of design choices, where transport-focused pricing policies can raise the probability of progressive outcomes by nearly 48 percentage points \citep{ohlendorf2021distributional}. Together, these statistics not only quantify effectiveness but also provide guidance for stricter, more equitable, and more Paris-aligned policy design.

\section{Discussion}

This umbrella review confirms a fundamental divergence in the evidence on the effectiveness of carbon-based climate policy instruments. On the one hand, compliance mechanisms such as carbon taxes and ETS appear capable of producing measurable and significant emissions reductions, though their real-world impact is critically conditioned by policy design---particularly the stringency of price levels and the scarcity of allowances. The moderate certainty of this conclusion under GRADE suggests that policymakers can reasonably rely on well-designed carbon pricing to contribute to mitigation goals. The main caveat is not intrinsic efficacy but political feasibility and distributional impacts, which require careful revenue-use design to secure equity and public acceptance, as documented by \citet{ohlendorf2021distributional} and \citet{klenert2018making}.

On the other hand, the evidence for voluntary carbon markets is substantially more pessimistic. High-certainty evidence pointing to systemic and large overestimation of offset climate benefits constitutes an alarming conclusion. Additionality, leakage, and permanence challenges are not merely technical imperfections but structural failures that undermine the fundamental premise of these markets---that one ton of CO\(_2\) emitted can be \lq neutralized\rq by one ton reduced elsewhere. As argued by \citet{romm2025are} and \citet{probst2024systematic}, the persistence of these problems despite decades of reform attempts suggests they may be inherent to voluntary, project-based crediting models. Corporate demand for cheap credits \citep{trencher2024demand} further exacerbates the issue, generating a \lq lemons\rq dynamic in which low-quality credits displace higher-quality ones. This has major implications for corporate net-zero strategies that rely heavily on offsets to substantiate neutrality claims.

The findings on social impacts of compensation projects, especially REDD+, add another layer of concern. Qualitative reviews such as \citet{bayrak2016ten} and \citet{dugasseh2024noncarbon} document patterns of dispossession and marginalization of local and Indigenous communities. This creates a fundamental tension: even if a project were to overcome technical carbon-accounting obstacles, it may still do so at an unacceptable social cost. The rhetoric of \lq co-benefits\rq (biodiversity, local development) is frequently invoked to justify these projects, yet the evidence suggests such benefits often fail to materialize or are unevenly distributed.

One limitation of this umbrella review is its dependence on the quality and scope of existing systematic reviews. Although we made efforts to appraise methodological rigor, biases in primary studies inevitably propagate upward through synthesis layers. For example, most ETS impact studies focus on Europe, and many REDD+ evaluations concentrate on the Amazon, limiting generalizability to other geographical contexts. Nonetheless, the consistency of findings across methodologically diverse reviews strengthens confidence in the main conclusions.

\section{Conclusion}

Aggregated empirical evidence synthesized through this umbrella review provides a clear yet nuanced picture of the role of carbon instruments in climate change mitigation. Carbon taxes and emissions trading systems appear to be effective tools, but their real-world impact depends critically on ambitious and equitable policy design. Prices must be high enough to drive deep decarbonization, and revenues must be managed to protect vulnerable groups and foster public acceptance. The key policy question is therefore not whether carbon pricing works, but how to make it work better.

By contrast, voluntary carbon markets, in their current form, are not a reliable mitigation instrument. Integrity problems are so deep and persistent that claims that these credits offset emissions are, in most cases, untenable. Relying on them to reach Paris-aligned targets is a high-risk strategy that could lock in weaker technologies and delay genuine climate action. Fundamental reform is needed, potentially including much stricter government regulation, a shift from \lq offsetting\rq to \lq contribution\rq claims without neutrality assertions, and a near-exclusive focus on durable, verifiable carbon removal rather than counterfactual-based avoided emissions. Without such transformation, the VCM risks doing more harm than good. Climate policy must be grounded in robust evidence, and the current evidence indicates that while some carbon instruments are part of the solution, others in their present state are part of the problem.

\end{document}